# Emergence and reconfiguration of modular structure for synaptic neural networks during continual familiarity detection

Shi Gu[a,b*], Marcelo G Mattar[c], Huajin Tang[d,e], Gang Pan[d,e]

[a] School of Computer Science and Engineering, University of Electronic Science and Technology of China, Chengdu, China
[b] Shenzhen Institute for Advanced Study, University of Electronic Science and Technology of China, Shenzhen, China
[c] Psychology Department, New York University
[d] College of Computer Science and Technology, Zhejiang University, Hangzhou, China
[e] State Key Laboratory of Brain Machine Intelligence, Zhejiang University, Hangzhou, China
To whom correspondence should be addressed: gus@uestc.edu.cn

## Abstract

While advances in artificial intelligence and neuroscience have enabled the emergence of neural networks capable of learning a wide variety of tasks, our understanding of the temporal dynamics of these networks remains limited. Here, we study the temporal dynamics during learning of Hebbian Feedforward (HebbFF) neural networks in tasks of continual familiarity detection. Drawing inspiration from the field of network neuroscience, we examine the network's dynamic reconfiguration, focusing on how network modules evolve throughout learning. Through a comprehensive assessment involving metrics like network accuracy, modular flexibility, and distribution entropy across diverse learning modes, our approach reveals various previously unknown patterns of network reconfiguration. In particular, we find that the emergence of network modularity is a salient predictor of performance, and that modularization strengthens with increasing flexibility throughout learning. These insights not only elucidate the nuanced interplay of network modularity, accuracy, and learning dynamics but also bridge our understanding of learning in artificial and biological realms.

## Introduction

Neural networks, both biological and artificial, are powerful computational models capable of learning and adapting to new information. Such models are used routinely in neuroscience and artificial intelligence (AI) for various applications related to learning. Neuroscience has long studied biological neural networks to unveil the principles of brains through understanding its organizational mechanisms and modeling the cognitive process[1], offering a wealth of information that can significantly enhance the AI model design and functionalities [2–4]AI, on the other hand, deploys artificial neural networks (ANNs) in a variety of applications, from computer vision and natural language processing, to more complex tasks that involve decision-making and prediction[5], holding the potential to decipher the brain's intricate mysteries [6,7]. While ANNs are often viewed as a digital manifestation of the brain's workings, a holistic framework that perceives an ANN as a dynamic neural system is conspicuously absent.

Current applications of AI in neuroscience can be broadly classified into two categories. The first category applies ANNs as prediction tools to strengthen the power of identifying associations, *e.g.*, utilizing sophisticated models to map brain connectome to labels and developing encoding models based on ANN features [6,8]. The second category builds recurrent neural network (RNN) models to execute cognitive tasks, targeting on understanding the relationship among tasks and the principles of cognition through manipulating the trained ANNs [9,10]. The methodological philosophy behind these two categories of approaches is that the similarity in performance may suggest similarity in the structure and representation. Yet, the dynamic nature of learning, intrinsic to both artificial and biological entities, remains largely uncharted, underscoring the chasms that persist in our comprehension of ANN methodologies and neural dynamics.

A substantial amount of research has been devoted to understanding the efficacy of ANNs, based primarily on concepts from computational optimization and statistical learning theory. Such perspectives, however, are insufficient to capture the dynamic, non-linear nature of learning in biological neural networks [11,12]. Indeed, while we are now able to construct networks capable of impressive feats[13], grasping their underlying learning dynamics is still an emerging frontier. This calls for versatile analytical instruments, capable of dissecting the temporal intricacies of ANNs, reflecting the persistent adaptability evident in biological neural networks during learning phases. We argue that a perspective grounded in computational neuroscience is indispensable, drawing from a set of techniques that we label *Artificial Network Neuroscience.*

Here, we illustrate these ideas by studying the learning dynamics of synaptic networks – specifically, Hebbian Feedforward (HebbFF) neural networks – during a continual familiarity detection task[14]. We choose this setting for two reasons. First, memory-related tasks have been widely studied in network neuroscience especially for the network reconfiguration across the learning procedure [15–18]. Second, the HebbFF model endowed with synaptic plasticity reproduces experimental results in the memory domain[14,19,20], making it an appropriate model for dynamic examination. We hypothesize that leveraging analytical techniques from network neuroscience will unearth profound insights into the dynamic shifts underlying neural network reconfigurations.

To study the learning dynamics of HebbFF, we developed a multi-pronged approach examining the dynamic reconfiguration of the networks from different perspectives. We begin with an analysis of the modularity over temporal scales and its relationship to variations in task accuracy and distribution entropy across diverse learning paradigms. We then explored the synchronicity of states during the training phase and its subsequent correlation with accuracy. We show that network modularization enhances with learning, and that network flexibility serves as a robust metric encapsulating model performance, in line with results from neuroscience in biological organisms. We hope that our findings will shed light on the interplay between network modularity, accuracy, and learning dynamics, and ultimately advance our understanding of artificial neural networks and their biological counterparts.

## Results

We explored the dynamic reconfiguration of artificial neural networks, focusing on a class of recurrent (RNNs) called Hebbian Feed Forward Network (HebbFF). This network determines the familiarity of a stimulus $x(t)$ based on whether it matches a stimulus encountered at a prior time-step $x(t-R)$[14]. The network input is an $N$-dimensional vector $x(t)$ and the output $y(t)$ indicates whether $x(t)$ equals $x(t-R)$ (Figure 1a). Note that this task setting resembles classical working memory paradigms, which have been investigated widely in network neuroscience terms of dynamic network reconfiguration[15,17]. The parameters of this continual familiarity detection task include a repeat interval length, $R$, and a vector length, $N$, which we set to $R = 5$ and $N = 100$. We use a neural network with 120 units in the hidden layer to provide sufficient representation power for encoding the input (Figure 1b). A more detailed discussion on the memory capacity can be found at [14].

We trained 120 network instances representing different subjects. Each network was trained on a training dataset of 2500 samples for 1000 epochs. After every 10 epochs, we evaluated the partially trained model on a testing dataset comprised of a separate 2500 samples, examining the activation of the hidden layer units to investigate the modularization effect. This analysis was performed on a $120 \times 2500$ matrix with 120 regions (units) and 2500 time points. As the HebbFF updates the plasticity matrix through the task, the activation vectors contain information about short memory, which support the combination of neighboring cases into time-windows. The sequence of 2500 time points was divided into 50 time-windows each with length 50. The dynamic "functional" connectivity matrices [21] were then constructed as the Pearson's matrix within each of the 50 time-windows (Figure 1c).

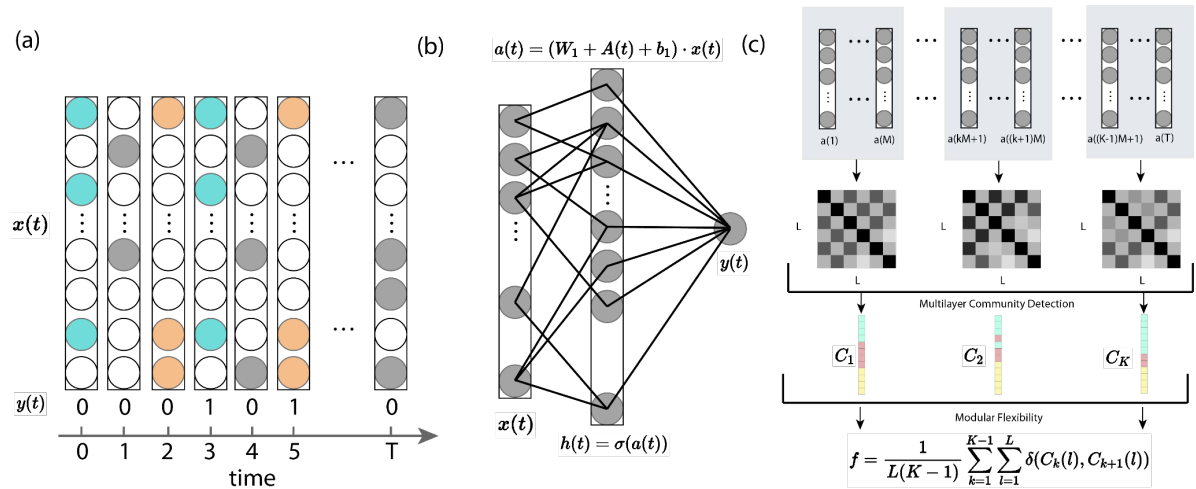

**Figure 1. Dynamic Reconfiguration and Learning Patterns in HebbFF Network.** An illustration of the continual familiarity detection task. The network output indicates if a given stimulus $x(t)$ matches the stimulus presented $R$ time-steps ago, $x(t-R)$. Here, R=3. (b) The HebbFF network, utilizing a hidden layer, encodes the input stimulus and predicts its familiarity through a linear classifier. (c) Analogous to the brain network, hidden layer activations are extracted in response to a task sequence, enabling the construction of multilayer functional networks through correlation within time windows. Here $M$ is the length of time window, and $K$ is the number of time windows. These multilayer structures not only illuminate the evolving modular dynamics over time but also echo the observations on modular flexibility and network entropy seen in subsequent figures.

In cognitive neuroscience, modularization is recognized as a pivotal mechanism that augments both computational efficiency and flexibility by facilitating localized information processing [22]. To understand the modularization process in the HebbFF learning, we analyzed the evolution of the network's community structure throughout the learning trajectory. We found that there was a pronounced decline in the number of communities early in the learning trajectory, concomitant with an escalation in overall network modularity (Figure 2 a,b). Such trends are indicative of a scenario wherein the quantitative reduction in module count is counterbalanced by an amplification in the representational capability of the extant modules. A deeper probe into the modular allegiance matrix across varied learning epochs reveals a nascent modular structure as early as epoch 10 (Figure 2c). By epoch 200, this structure began to delineate with more pronounced modules. By epochs 390 and 580, we observed a further crystallization of these modules, with distinct community structure surfacing. By epochs 770 and 960, the matrix displayed pronounced and contrastive modules, underscoring a heightened modular allegiance. Such dynamics underscore the iterative refinement and consolidation of a community structure during the learning process, echoing the behavior of biological neural networks. These results highlight the emergence of modular, localized processing in fine-tuning learning dynamics, raising the important question of whether such structure is related to task performance.

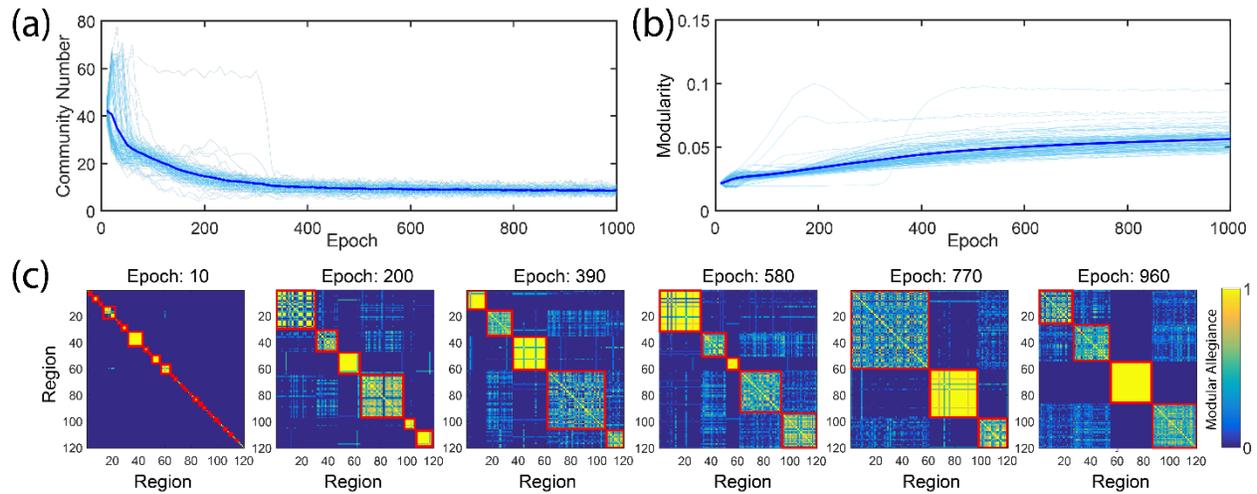

**Figure 2. Modularization of hidden layer activation over training.** (a) The number of communities decrease and (b) the modularity increases through the training. (c) The modularization becomes more significant through trainings with less modules and more contrastive modules.

To analyze more closely the modularity increase during learning, we examined the activations of the hidden layer. We found that, on average, the hidden layer's activation initially increased its very early stages, only to decline as training matured towards its latter phases (Figure 3a). Such a reversal could result from the bolstering of the distribution's negative spectrum or a possible attenuation of its positive counterpart. Additional insights into this behavior can be extracted from observing the activation distributions across model instances across select epochs, from early to late stages (Figure 3e). We found that the probability of near-zero activations increased throughout learning, consistent with the increased modularity described previously (Figure 2c).

We also found interesting patterns in the kurtosis and skewness of the activation distributions. We observed a slight contraction in the kurtosis of hidden layer neural activations early in learning, followed by a drastic expansion in the later stages (Figure 3c). This trend corroborates our observations in Figure 3e, which demonstrate an accumulation of activation values in the vicinity of zero. The skewness, meanwhile, displayed an intriguing pattern. Early in learning, the distribution of neural activations showed a consistent decline in skewness, representing an increase in the symmetry of the distribution. Yet, as training enters its intermediate and late phases, the skewness bifurcated into dual trajectories. While one set of networks followed a steady increase in skewness, the other showed a rapid increase before a slower decrease. This raises the question of whether these distinct trajectories represent distinct modes of learning, and whether they are relevant for behavior .

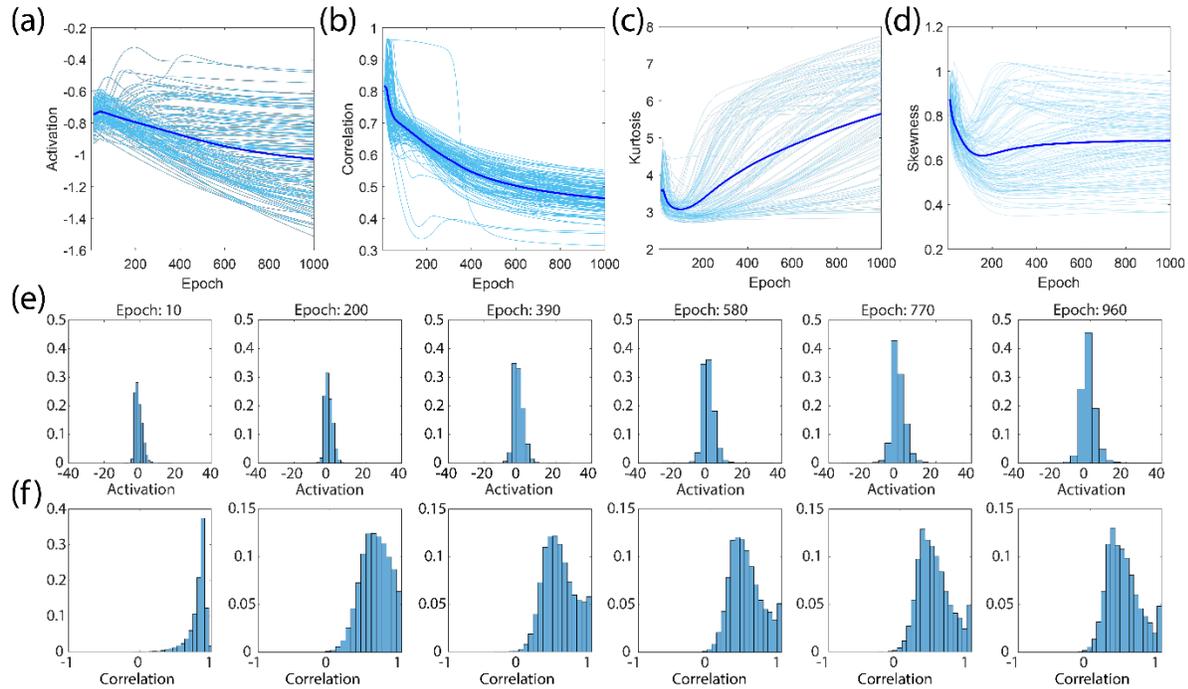

**Figure 3. Evolution of Activation Patterns in the Hidden Layer Over Training Periods.** (a) Tracing the trajectory of average activation in the hidden layer, an initial moderate increase is evident in the early training epochs, succeeded by a subsequent decline persisting until the latter epochs. (b) An analysis of the activation values' correlation among neurons in the hidden layer exhibits a consistent decrease as training progresses. (c) A study of the kurtosis of the activation values across epochs shows an initial decrease, followed by a resurgence in the later stages of training. (d) The skewness of activation values delineates an intriguing pattern: an initial decline, transitioning into a bifurcation in later epochs, manifesting both an ascendant trajectory and an alternate trajectory with an initial surge followed by a decline. (e) Epoch-specific histograms depict the activation distribution across select epochs, illustrating a growing concentration of values around zero and a reduction in distributional asymmetry as training advances. (f) Histograms across epochs, analogous to those in (e), elucidate the shifts in correlation values. Early training predominantly showcases correlations proximate to unity. As training matures, the distribution gravitates towards zero, concluding in a mildly zero-skewed distribution punctuated by an isolated peak at one.

To gain additional insights, we examined the correlations within the hidden layer. We found that correlations tended to decrease throughout learning (Figure 3b). In early stages, hidden layer activations were highly correlated (Figure 3f, left). As training unfolded, however, the distribution of correlation shifted towards zero, culminating in a slightly zero-skewed distribution punctuated by an isolated peak at one (Figure 3f, right). Such behavior suggests a de-synchronization within the network, suggesting an augmentation in representational power that may underlie an increase in model performance.

In network neuroscience, a modular structure holds dual significance: firstly, as a descriptor of learning phase differentiation; secondly, as an indicator of inter-individual variability during cognitive assessments. To understand the interplay between modularity and performance within HebbFF, we explored the correlation dynamics between modularity and accuracy, both within specified epoch windows and across disparate model instances (Figure 4).

Looking across windows of 100 epochs, we found a consistently positive correlation between modularity and accuracy, suggesting that more modular networks tend to perform better. This correlation approached one during the median training stages, subsequently retracting to approximately 0.6 in late stages (Figure 4a). This trend is congruent with the overarching rise in modularity as training progresses depicted previously (Figure 2b). Given the near-monotonic increase in modularity (Figure 2b) and in task accuracy (Figure 4a), the trends in correlations suggest that the heightened modularity facilitate, in median training stages, the memorization of a broader set of states.

Holding epochs constant and examining correlations across model instances, we found a distinct pattern. The modularity-accuracy correlation demonstrates an initial upswing during the formative training epochs, which then descends to negative magnitudes during the intermediate and concluding phases (Figure 4c). Examining these patterns across epoch windows and model instances reveals a particularly intriguing trend (Figure 4d). While a single model's learning curve exhibits a positive correlation between accuracy and modularity, this association switches to negative when examined over model instances. These findings suggest that, within models subjected to extended training, a robust modular structure — manifesting as markedly modularized activation states — could potentially compromise representational capacity, consequently attenuating overall performance.

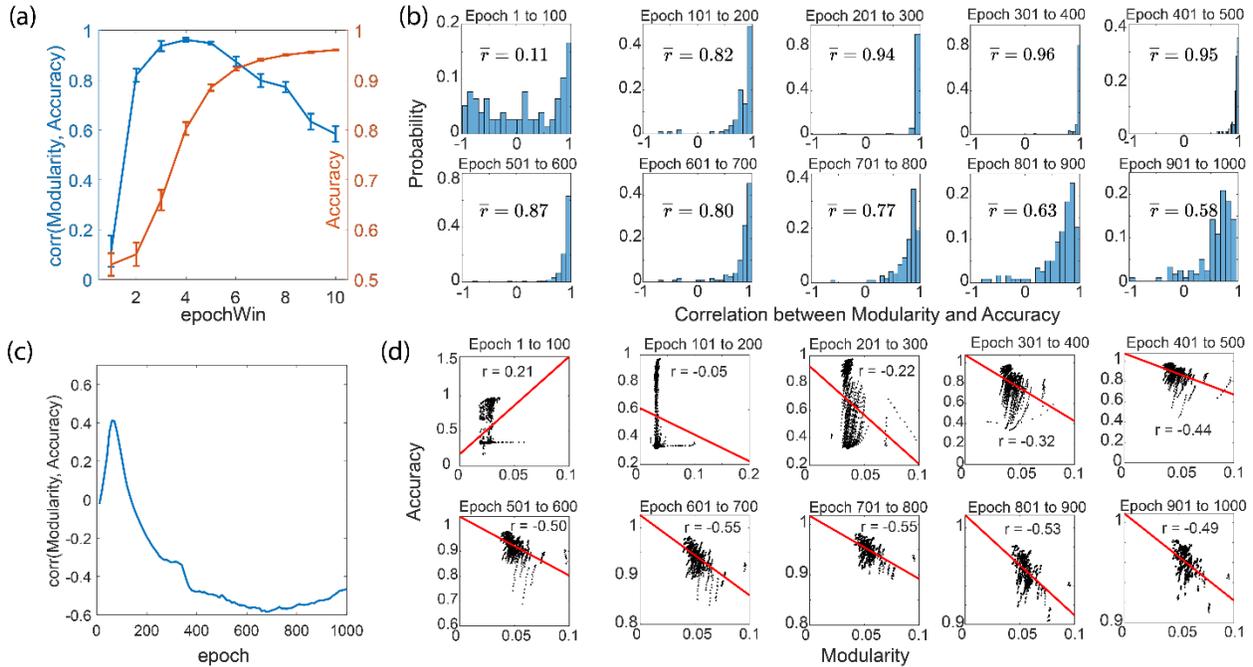

**Figure 4. Correlation Dynamics between Network Modularity and Performance Accuracy.** (a) Correlation progression between modularity and accuracy across distinct epoch windows, each comprising 100 epochs. Initial stages showcase a surge in correlation nearing unity, which subsequently declines to approximately 0.6 by the concluding stages. (b) Depiction of correlation distributions corresponding to distinct epoch windows, capturing the variability within individual error bars of panel (a). (c) Overall correlation trajectory between modularity and accuracy, calculated across diverse model instances. A marked positive correlation during the formative training stages inversely transitions to a negative domain during intermediate and advanced stages. (d) Comprehensive scatter plots showcasing modularity-accuracy correlation trends for specified epoch windows, aggregated across different model instances. These reveal the intricate balance between modular architecture and performance, particularly in extensively trained networks.

The results thus far illustrate the relevance of network modularity, a static metric of mesoscale structure. To complement these findings, we examined also the network's, modular flexibility, which quantifies the dynamic reconfiguration rate of the networks' modular structure. In biological networks, modular flexibility has been empirically linked to task execution proficiency in neuroscience studies, displaying distinct patterns across learning phases. Here, we examined modular flexibility to assess its connection with model performance across learning stages and across model instances (Figure 5).

To examine the dynamics of modular flexibility, we applied the multi-slice Louvain algorithm [23] to the 50-layer networks to obtain the temporally varying community association. We then calculated the network flexibility [15] for each model instance. Initially, we segmented the entire learning procedure into 10 epoch windows. Within each of these windows, we calculated the correlation between flexibility and accuracy, and subsequently aggregated these correlations to produce a composite histogram and error bars. We found a remarkable parallel between increase in modular flexibility and accuracy over the learning process (Figure 5a). Similarly to the trend found for modularity, correlations between flexibility and accuracy (across model instances) increased in the early learning stages before declining in the subsequent stages (Figure 5b). Interestingly, within trained models, we found that lower flexibility —
which we associated with increased representational stability — may predict higher memory
performance. We found several correlation peaks in epochs 101-200 and 201-300, resonating with the
latter phases of the warm-up and the onset of the progressive stage, respectively. Epoch-specific
correlations for the 120 model instances display an intriguing pattern (Figure 5c). Although flexibility and
modularity follow similar trajectories, the peak of the correlation between flexibility and accuracy
precedes the peak of the correlation between modularity and accuracy. This suggests that dynamic metrics
(e.g., flexibility) might be precursors to the emergent modular representations throughout training.
Furthermore, while flexibility contributes to learning, optimal performance may require stable
representations.

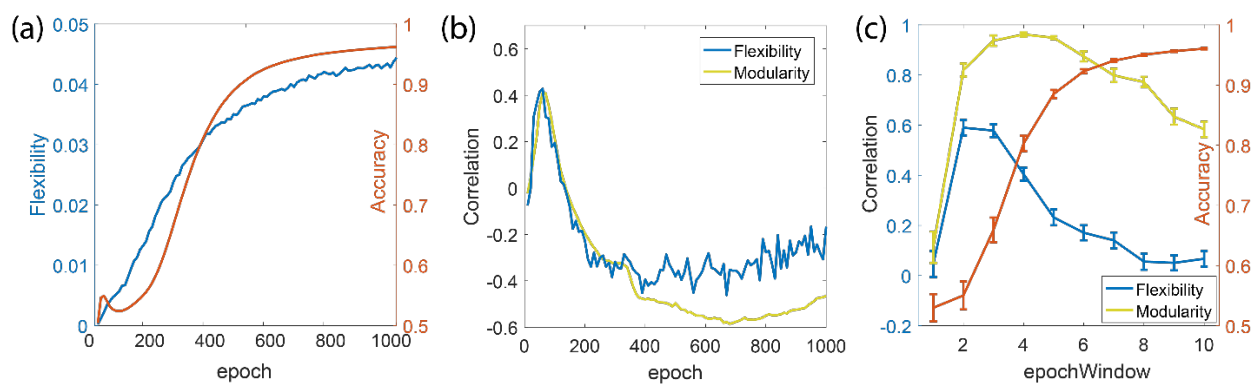

**Figure 5. Evaluating the interplay between accuracy and modular flexibility during training.** (a)
Over the training progression, modular flexibility consistently elevates in tandem with accuracy. Initially,
there's a distinctive oscillation in the accuracy curve, mirroring a "warm-up" phase, while modular
flexibility follows a clear upward trend. (b) The epoch-wise correlation between modular flexibility and
accuracy (spanning individual model instances) first ascends, recording positive values, before
descending into negative territory. (c) Analyzing correlations within designated epoch windows, we
discern that while both modularity and flexibility exhibit an upward trajectory, flexibility achieves its
correlation zenith earlier, subsequently declining to near-zero levels in the concluding stages.

The patterns observed in the distributions of hidden unit activations (Figure 3a-d), when viewed alongside
the evolution in modular flexibility (Figure 5a), suggest varying learning trajectories across different
model instances. To discern whether prominent learning patterns exist, we analyzed the performance of
HebbFF on the test set throughout the learning process (Figure 6). This analysis revealed two distinct
learning modes. In the first mode (Learning mode I), accuracy remained nearly constant for the first 200
epochs (Figure 6a,b). In the second mode (Learning mode II), accuracy increased rapidly in the first few
epochs before decreasing again around epoch 100 (Figure 6e,f).

In brain networks, network flexibility represents the community adaptation rate throughout time, with
higher flexibility often correlating with heightened cognitive task performance. We wished to determine if
this relationship between network flexibility and performance is also found in HebbFF networks.
Interestingly, both learning modes, despite differing warm-up behaviors, consistently exhibited a
monotonic increase in modular flexibility, possibly to increase the network's representation capability and accuracy (Figures 6c,g). This raises the possibility that modular flexibility is a potentially broader network characteristic, transcending its traditional heuristic function.

To further examine the representational capacity of the hidden layers, we gauged the distribution entropy of hidden layer activation, interpreting it as the average informational content. For mode I, despite the averaged entropy curve increasing before the 100-epoch mark, significant fluctuations were noted between 100 to 300 epochs (Figure 6d). In contrast, mode II experienced a nearly monotonic rise in entropy after a transient in the first few epochs, suggesting consistent informational augmentation parallel to increasing modular flexibility. Intriguingly, both modes demonstrated entropy [24] growth in latter stages, implying that global optima optimization correlates with enhanced representation.

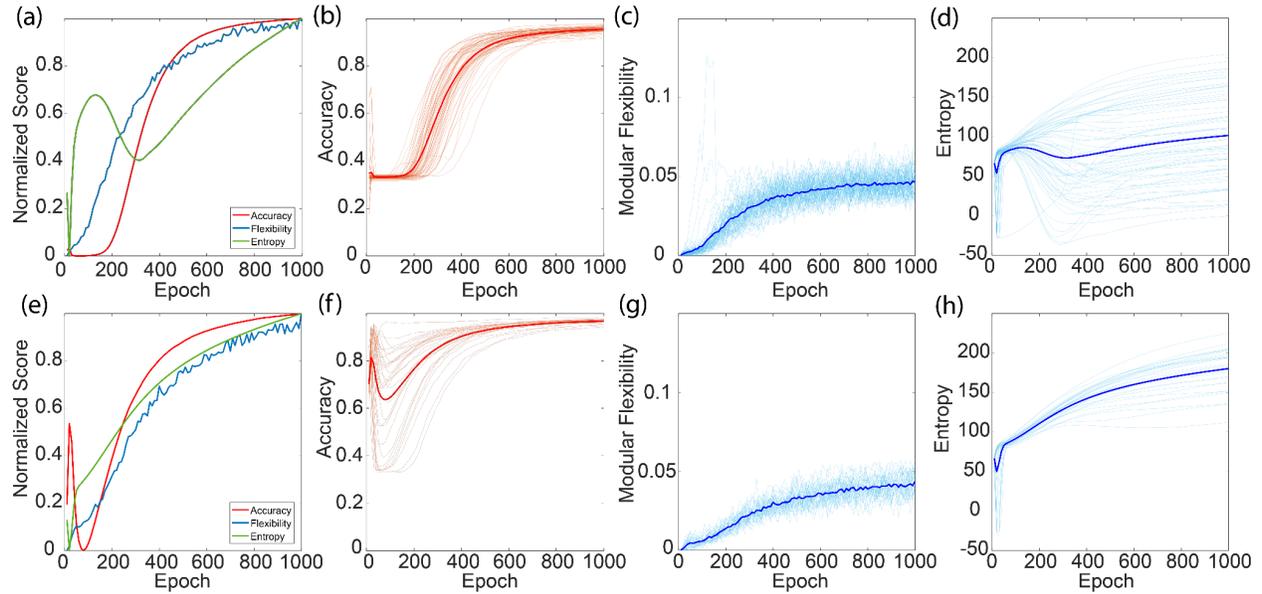

**Figure 6. Divergent Modes of Learning in HebbFF Networks.** (a) Learning mode I showcases different evolutionary trends for accuracy, modular flexibility, and distribution entropy. The accuracy curve (b) exhibits a warm-up period of approximately 20 epochs before ascending to its upper limit. Simultaneously, the modular flexibility (c) consistently grows, reaching a plateau, while the distribution entropy of the hidden layer (d) peaks during the warm-up phase, then diminishes during the early accuracy surge before experiencing another incline as training concludes. In contrast, Learning mode II (e) reveals distinct patterns in accuracy, modular flexibility, and distribution entropy compared to mode I. Specifically, the accuracy (f) swiftly climbs to a sub-optimum, then recedes, facilitating further exploration within the loss landscape to pinpoint global optima. In spite of the fluctuating accuracy, both the modular flexibility (g) and distribution entropy (h) perpetually ascend. For clarity, the normalized score represents the original measurement adjusted to fit within a 0 to 1 range.

## Discussion

In our study, we studied the learning dynamics of the Hebbian feed forward neural network, identifying the existence of two distinct learning modes and their associated patterns of modularization. These findings underscore the potential of network neuroscience to characterize the learning behaviors of artificial neural networks. By examining the evolution of modular flexibility, modularity, and entropy throughout the learning process, we obtained significant insights into the learning behavior of Hebbian feed forward neural network in the familiarity detection task. Notably, the relationship between flexibility and accuracy provides compelling evidence for the relevance of network methods for the analysis of artificial neural networks.

Prior studies consistently demonstrate a correlation between network segregation and integration with the execution of cognitive tasks [25–27]. Typically, results are interpreted in terms of challenging tasks requiring integration between modules, with the need for integration decreasing through learning [25]. Consistent with prior research, we found that the modularity of HebbFF networks increased during the learning process. These changes reflected the enhancement of modularity values associated with learning [28] and neurodevelopment [29]. The fluctuations in flexibility throughout the learning process also mimic prior findings [15,16], underscoring the role of increased flexibility in supporting task learning for given subjects. In contrast, we observed a negative association between modularity and accuracy, as well as between flexibility and accuracy, in late learning stages. This aligns with the decrease of modularity[30] and flexibility[31] through the adolescent neurodevelopment.

In our investigation of HebbFF networks, we also observed that the processes of segregation and integration occurred not only among different functional modules, but also within the representation of diverse features. This suggests that network modularization, as it pertains to feature representation, may hold significant relevance for understanding the functionality of the brain. In the brain, different systems handle different aspects of the input and decision-making processes, implying that features could be encoded across multiple interconnected systems. This holistic approach to understanding neural networks, by considering the role of feature representation in network modularization, may provide unique insights into the complex dynamics of both artificial and biological learning systems. The consistency of the results on HebbFF and brain networks support the link between the network characteristics and the cognitive execution [32]. In addition, the role of modularization in the representation is also supported in previous work varying from neuroscience [33] to machine learning [34]. Thus, beyond offering a nuanced interpretation of network methodologies, our study points to a potentially unifying understanding of dynamic reconfiguration roles in both human cognition and machine learning.

Existing network neuroscience analyses of functional brain networks typically model the brain as a dynamic system that either follows a random walk [35] or a given geometric flow [36]. These analyses are based on the implicit assumption that the variation of the information segregation and integration would cause corresponding difference in the signal space. Here, since functional connectivity was constructed directly from patterns of activation in the hidden layer units, we provide a more directive examination on the relationship between the learning behavior and the representation topology. This approach offers numerous insights into the performance of ANNs, including how the topology of the feature representation varies in early stages of learning, even when accuracy does not increase. While previous work [12] investigates how network topology affects the performance, our work adds to this literature by suggesting that, in addition to the network structure and its associated static measurement, the networked dynamics may provide additional evidence on how the model performance is related to the structure. This aligns with neuroscience findings that the dynamics of functional connectivity provides better predictions of cognitive scores than static measures [37]. Accordingly, out work suggests a new family of measurements and can facilitate the design of brain-inspired neural networks.

More broadly, our work contributes to the field of *AI for Neuroscience* by providing a better understanding of the dynamical behavior of artificial neural networks through a neuroscience-inspired lens. The endeavor to understand the human brain and develop efficient artificial intelligence systems has often been a mutually beneficial process. However, a major challenge in *AI for Neuroscience* has been bridging the gap between the static, linear analysis commonly used in machine learning and the dynamic, non-linear characteristics that are intrinsic to the biological brain. Traditional methods for analyzing neural networks, such as studying the weight and bias parameters, may not capture the complete picture of how an artificial neural network learns, adapts, and evolves over time. In this context, our study provides new tools and methodologies to better understand these dynamics. By constructing *brain-like* networks from HebbFF and applying techniques like community detection and modular analysis, we mirror the modular structure and dynamic reconfiguration seen in the biological brain. In essence, we provide an avenue to study the temporal evolution and adaptation of artificial neural networks during training, similar to the continual reconfiguration observed in the brain during learning. This approach enhances our understanding of the similarities and differences between artificial neural networks and the human brain, opening new avenues for improving the design and training of artificial neural networks. Our work also takes a step towards closing the gap between the simplistic activation and loss landscapes usually employed in AI research and the complex, high-dimensional, and dynamic landscapes that are likely in the brain. By investigating the synchronization and entropy within the HebbFF network, we are not just considering the performance of the AI system but also the underlying system dynamics that facilitate learning.

In sum, our discoveries contribute significantly to the broader aim of intersection between AI and Neuroscience - using AI not only to replicate but also to understand and learn from the intricate workings of the brain. The tools and methods we developed present new opportunities to study learning dynamics in both artificial and biological neural networks. Such cross-fertilization of ideas can potentially lead to more efficient, adaptable, and robust AI systems while providing insights into the neuroscience of learning and memory. We note that our research so far focused on memory tasks, which are well-suited for Hebbian feedforward networks that are inherently amenable to descriptions based on modularization. Future work should expand our approach to other cognitive tasks like multimodal matching, value decision, and perception tasks, as well as to other types of ANNs including RNN and deep feedforward networks. We hope that multimodal continual tasks learned through complex networks could serve as a digital analogue of the brain in terms of cognitive execution and may provide novel insights into how functional modules reconfigure to support complex tasks.

## Method

**Hebbian Feedforward Network Architecture.** The HebbFF Network takes an $N \times 1$ vector $\boldsymbol{x}(t)$ and returns a $y(t)$ that indicates whether $\boldsymbol{x}(t) = \boldsymbol{x}(t - R)$. Here $R$ is the repeat interval length. In this work, we set $R = 5$ and $N = 100$. The HebbFF Network consists of three layers, the input, output, and hidden layers. The hidden layer consists of $M$ neurons. We use an $M \times 1$ vector $\boldsymbol{a}(t)$ to denote the hidden state and $\boldsymbol{h}(t) = \sigma(\boldsymbol{a}(t))$ to denote the activation state after an activation function $\sigma(\cdot)$, where $\sigma(x) =$ $\frac{1}{1+\exp(-x)}$ takes the form of the sigmoid function. The hidden state $a(t)$ is obtained through a linear transformation on the input $x(t)$, which is given by

$$a(t) = (W_1 + A(t))x(t) + b_1, \quad (1)$$

where $W_1$ is an $M \times N$ matrix denoting the affine transformation and $b_1$ is an $M \times 1$ vector denoting the bias. The matrix $A(t)$ is the plasticity matrix that is updated at every time step to take care of the memory. It is calculated as

$$A(t+1) = \lambda \cdot A(t) + \eta \cdot h(t)x(t)^T, \quad (2)$$

where $\lambda$ is the learnable decay parameter and $\eta$ is the learnable familiarity learning rate. When $\eta > 0$, it is called the Hebbian learning; when $\eta < 0$, it is called the anti-Hebbian learning. As demonstrated in [14] and supported by our own experiments, we adopt the anti-Hebbian learning rules for the familiarity detection. Further, the readout is given by

$$\hat{y}(t) = \sigma(W_2 h(t) + b_2), \quad (3)$$

where $W_2$ of dimension $2 \times M$ and $b_2$ of dimension $2 \times 1$ are learnable transformation weight and bias. The training loss is then set as

$$L = \frac{1}{T} \sum_{t=1}^{T} y(t)\log(\hat{y}(t)) + (1 - y(t))\log(1 - \hat{y}(t)), \quad (4)$$

which represents the binary cross-entropy between the label $y$ and prediction $\hat{y}$. The activation values $a(t)$ and $A(t)$ are updated through the training thus can be used to analyze the network dynamics. The network structure is shown in Figure 1 (a) and Figure 1 (b).

**Construction of Brain-like Networks from the HebbFF Activation.** For an instance of HebbFF, after being training for $t$ epochs, we evaluate the model on the testing dataset $X_{test}$ of dimension $N \times T$ and collect the hidden states into a matrix $HA^t = [a(1), ..., a(T)]$ of dimension $M \times T$, which can be taken as the recorded activation sequence of the HebbFF for continuously executing $T$ tasks. We then divide the full $HA^t$ into $L$ time windows each of the length $T/L$. As the length parameter $T$ is fully controllable, we assume that $T$ can be divided up by $L$ for simplicity. We obtain a multilayer network $\{A^t_{ijl}\}$ where $A^t_{ijl}$ is the Pearson's correlation value of the $i^{th}$ and $j^{th}$ rows in $HA^t$ within the $l$-th time window. Based on these $\{A^t_{ijl}\}$ along the full $K$ epochs' training, we can perform network-based analysis to characterize the learning behavior of HebbFF through training.

**Modularization.** Community detection is a method that decomposes a system into subsystems [23]. For a given multilayer network $\{A_{ijl}\}$ where $i$ and $j$ denote the regions and $l$ denote the layers. The multilayer modularity function is given as

$$Q = \frac{1}{2\mu} \sum_{ijlr} \left( A_{ijl} - \gamma_l \frac{k_{il}k_{jl}}{2m_l} \delta_{lr} + \delta_{ij} C_{jlr} \right) \delta(g_{il}, g_{jr}), \quad (5)$$

where the adjacency matrix of layer $l$ has component $A_{ijl}$, $\gamma_l$ is the resolution parameter of layer $l$, $g_{il}$ and $g_{jr}$ give the community assignments of nodes $i, j$ in layers $l, r$, $k_{il}$ is the strength of node $i$ in the layer $l$ with $2\mu = \sum_{jr} \kappa_{jr}$, $\kappa_{jl} = k_{jl} + c_{jl}$, and $c_{jl} = \sum_r C_{jlr}$. When $A_{ijl}$ is signed, one can split the positive and negative part and construct the modularity function similarly as shown in [38]. Through maximizing $Q$, we can get the community structure $g_{il}$ of each node in each layer, which allows us to further investigate the networked dynamics of modules.

**Dynamic Reconfiguration of HebbFF Networks.** Based on the constructed network series $\{A_{ijl}\}$ and their associated community structure $g_{il}$, we can then define the modular flexibility $f_i$ as the changing frequency of the community association over time [15], i.e. $f_i = \frac{1}{L-1}\sum_{l=1}^{L-1}\left(1 - \delta\left(g_{i,l}, g_{i,l+1}\right)\right)$. Further, we can define the modular flexibility of the system as $f = \frac{1}{M}\sum f_i$. If the system has a high flexibility, it indicates that the module structure of the system changes fast, thus suggesting a high flexibility in the representation of the learned features.

**Information Entropy of Activation Variables.** For a HebbFF Network, the modular structure of the activation-induced correlation network is supported by the similarity of neuron's activation patterns. In order to quantify the strength of different region's participation in support the temporal dynamics, we adopt the entropy of a random variable quantifies the average information contained in the outcome [24]. It is defined as the expectation of the log of the density for a continuous distribution. Mathematically, it is denoted as $H(x) = E_x[-\log p(x)]$. For each HebbFF, when it processed $K$ samples, we can collect $K$ activation vector $\boldsymbol{a}_1, ..., \boldsymbol{a}_K$. Based on these $\boldsymbol{a}_1, ..., \boldsymbol{a}_K$, we can estimate the $H(\boldsymbol{a})$ as the information contained by the hidden layer. Here, as our purpose is not accurately defining the information quantity, thus for simplicity, we assume that these $\boldsymbol{a}_i$ follows a multivariate gaussian distribution.

**Statistic Inference**
The p-value associated with the Pearson's correlation r is calculated using a t-distribution with $n-2$ degrees of freedom, where $t = \frac{r\sqrt{n-2}}{\sqrt{1-r^2}}$ and $n$ is the number of samples.

# Data & code availability statement
All original data in this work was generated by program provided at https://github.com/dtyulman/hebbff with some homemade adaption for storage and parallel execution. The code for data processing and analysis can be found at https://github.com/gushiapi/flex_hebbff. Any further information about the data and script is available upon request.

# Author Contribution
S.G. designed and led the research. S.G. developed and performed the analysis. S.G., M.G.M, H.J.T, and G.P. wrote the paper.

# Acknowledgement
S.G. is supported by NSFC Key Program 62236009, Shenzhen Fundamental Research Program (General Program) JCYJ 20210324140807019, NSFC General Program 61876032, and Key Laboratory of Data Intelligence and Cognitive Computing, Longhua District, Shenzhen.